\RequirePackage{amsmath}
\documentclass[epj]{svjour}
\usepackage{latexsym}
\usepackage{graphics}
\usepackage{appendix}
\usepackage{graphicx} % Include figure files
\usepackage{multirow}
\usepackage{footnote}
\usepackage{units}
\usepackage{epstopdf}
\usepackage{epsfig}
\usepackage{lpic}
\usepackage{color}
\usepackage[makeroom]{cancel}
\usepackage[english]{babel}
\usepackage{xfrac}
\newcommand{\tHe}{\ensuremath{{}^3\mathrm{He}}}

\newcommand{\Cs}{\ensuremath{{}^{133}\mathrm{Cs}}}
\newcommand{\Rb}{\ensuremath{{}^{87}\mathrm{Rb}}}
\begin{document}\sloppy
\title{PicoTesla absolute field readings with a hybrid \boldmath{$^3$}He/\boldmath{$^{87}$}Rb magnetometer}

\author{
Christopher Abel\inst{1}\textsuperscript{,}\thanks{e-mail: c.abel@sussex.ac.uk}
\and Georg Bison\inst{2}\textsuperscript{,}\thanks{e-mail: georg.bison@psi.ch}
\and W. Clark Griffith\inst{1}
\and Werner Heil\inst{3}
\and Klaus Kirch\inst{2,4}
\and Hans-Christian Koch\inst{2}
\and Bernhard Lauss\inst{2}
\and Alexander Mtchedlishvili\inst{2}
\and Martin Pototschnig\inst{4}
\and Philipp Schmidt-Wellenburg\inst{2}
\and Allard Schnabel\inst{5}
\and Duarte Vicente Pais\inst{2}
\and Jens Voigt\inst{5}
% etc
%\thanks{\emph{Present address:} LogrusData, Vienna, Austria}%
}                     % Do not remove
\institute{
Department of Physics and Astronomy, University of Sussex, Brighton, UK
\and Paul Scherrer Institute, CH-5232, Villigen, Switzerland
\and Institut f\"ur Physik, Johannes-Gutenberg-Universit\"at, D-55128 Mainz, Germany
\and Institute for Particle Physics and Astrophysics, ETH Zurich, 8093 Zurich, Switzerland
\and Physikalisch-Technische Bundesanstalt, Berlin, Germany
}
\date{Received: February 18, 2019}
\abstract{
%\PACS{{1}{NOT DONE YET} \and
%      {07.55.Ge}{Magnetometers for magnetic field measurements}\and
%      {32.30.Dx}{Magnetic resonance spectra}\and
%      {42.62.Fi}{Laser spectroscopy}} % end of PACS codes
%
We demonstrate the use of a hybrid \tHe{}/\Rb{} magnetometer to measure absolute magnetic fields in the pT range.
The measurements were undertaken by probing time-dependent \tHe{} magnetisation using \Rb{} zero-field magnetometers.
Measurements were taken to demonstrate the use of the magnetometer in cancelling residual fields within a magnetic shield. It was shown that the absolute field could be reduced to the \unit[10]{pT} level by using field readings from the magnetometer.
Furthermore, the hybrid magnetometer was shown to be applicable for the reduction of gradient fields by optimising the effective \tHe{} $T_2$ time.
This procedure represents a convenient and consistent way to provide a near zero magnetic field environment which can be potentially used as a base for generating desired magnetic field configurations for use in precision measurements. %, such as those required by the next-generation neutron electric dipole moment experiments.
} %end of abstract
\maketitle
%
%%%%%%%%%%%%%%%%%%%%%%%
\section{Introduction}
%%%%%%%%%%%%%%%%%%%%%%%
Many precision measurements require a thorough knowledge of surrounding magnetic fields, for example fundamental physics searches \cite{Safronova2018-af,Altarev2009-yf,Altarev2009-cc},
for electric dipole moments~\cite{Pendlebury2015-av,Abel2018-zb,Abel2018-qb,Graner2016-hy}, exotic spin coupling forces \cite{Dobrescu2006-uu,Afach2015-pt}, and axions~\cite{Abel2017-pw,Arvanitaki2014-gx,Graham2018-iv}, as well as general calibration of magnetic sensors \cite{Mohamadabadi2014-pn}.
It is generally difficult to produce an absolutely correct magnetic field with precise knowledge of both the direction and magnitude, especially due to the inevitable presence of small residual fields after typical degaussing procedures of magnetic shielding.
However, if one starts with an environment where the residual magnetic field is as close to zero as possible, any field can then in principle be generated on top of this using one or more calibrated coils.
\par
We present here a hybrid \Rb{}/\tHe{} magnetometer that can be used to provide absolute magnetic field readings based on the free spin precession (FSP) of \tHe{}, monitored by optical \Rb{} magnetometers that detect the small magnetic field variations generated by the precessing \tHe{} polarisation. 
This magnetometer can be used to zero a residual field environment, upon which a desired magnetic field can be applied using a separate coil system. \par
%
%This magnetometer can be used to establish a near zero magnetic field environment, upon which a desired magnetic field can be generated using a calibrated coil system.\par
This work follows on from a previously developed combined \tHe{}/\Cs{} magnetometer described in~\cite{Koch2015-vq}, capable of measuring magnetic fields around \unit[1]{$\mu$T} with sensitivities of \unit[$\sim$50]{fT} per \unit[180]{s} measurement cycle. 
In this work the \Cs{} readout sensors based on rf-driven magnetic resonance are replaced with \Rb{} sensors capable of operating at zero magnetic field. A similar concept was demonstrated by Cohen-Tannoudji et al. in 1969~\cite{Cohen-Tannoudji1969-tq}, where a low-field \tHe{}/\Rb{} magnetometer was used to measure an applied \unit[200]{pT} holding field. % and \tHe{} $T_2$ relaxation time~\cite{Cohen-Tannoudji1969-tq}.
Here we apply no explicit holding field but instead allow the \tHe{} to precess around the residual field inside a degaussed magnetic shield. After applying a field minimisation process with a set of correction coils, we have measured absolute fields below \unit[5]{pT}. We have also demonstrated minimisation of field gradients at the \tHe{} cell by maximising the effective $T_2$ relaxation time.

\par
\section{Measurement principle}
%%%%%%%%%%%%%%%%%%%%%%%%%%%%%%%%%%%%%%%%%%%%%%%%%%%%%%%%%%
% Steps of measurement. Polarising He, allowing to free precess and detection via the Rb.
Measuring low magnetic fields with a \tHe{} FSP magnetometer follows two steps; polarisation of \tHe{} in an enclosed volume, followed by detection of the free precession signal. \par
% Polarising the He using MEOP, no holding field as long precession times.
Metastability exchange optical pumping (MEOP) was used to polarise \tHe{} which has been shown to achieve polarisation levels up to $p \sim 80 \%$~\cite{Nacher1985-wo}.
At sufficiently low fields, when the Larmor frequency of the \tHe{} is much less than the polarisation pump rate, the helium can be polarised without the need for a holding field, such as that used in~\cite{Cohen-Tannoudji1969-tq}.
During MEOP, the \tHe{} polarisation is held in the direction of the pump laser propagation $\vec{k}_{He}$.
After polarisation, when the laser radiation is switched off, the \tHe{} polarisation precesses around the volume-averaged cell field vector.
% Free precession of the He around B_0
The frequency of the \tHe{} free-precession is free from light shifts commonly associated with optically pumped magnetometers~\cite{Kastler1963-jp}.
% Detection of this precession using the Rb magnetometer, what is the predicted value of the maximum precession?
The precessing magnetisation of the \tHe{} is then probed using zero-field \Rb{} magnetometers placed close to the \tHe{} cell. Using the Larmor frequency of the \tHe{}, the volume-averaged field magnitude at the centre of the cell can be calculated along with values for the relaxation time of the polarisation ($T_2$).
The high sensitivity of the \Rb{} zero field magnetometer to field changes of well below \unit[1]{pT} (AC below \unit[100]{Hz}) is used to precisely detect perturbations caused by the precessing \tHe{} magnetisation at the \Rb{} magnetometer's position.
%
%%%%%%%%%%%%%%%%%%%%%%%%%%%%%%%%%%%%%%%%%%%%%%%%%%%%%%%%%%%%%%
\section{Experimental apparatus} \label{sec:expApp}
%%%%%%%%%%%%%%%%%%%%%%%%%%%%%%%%%%%%%%%%%%%%%%%%%%%%%%%%%%
\begin{figure}
\centering
\includegraphics[width=0.9\columnwidth]{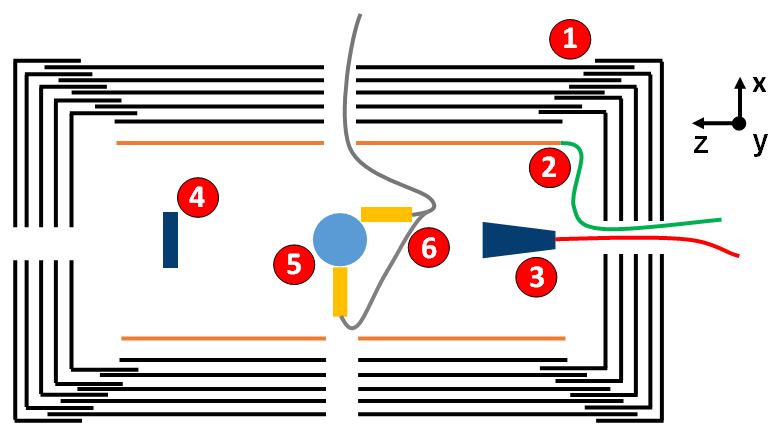}
\caption{\label{fig:apparatus}Schematic diagram of the hybrid magnetometer, along with the axis definitions. 1) magnetic shielding, 2) internal coils, 3) helium laser expansion optics, 4) \unit[1083]{nm} back-reflecting mirror, 5) \tHe{} cell, 6) \Rb{} magnetometers. The outermost shield layer has a length \unit[200]{cm} and a diameter \unit[94]{cm}. The innermost layer has length \unit[194]{cm} and diameter \unit[50]{cm}}
\end{figure}
\begin{figure}
\centering
\includegraphics[width=0.7\columnwidth]{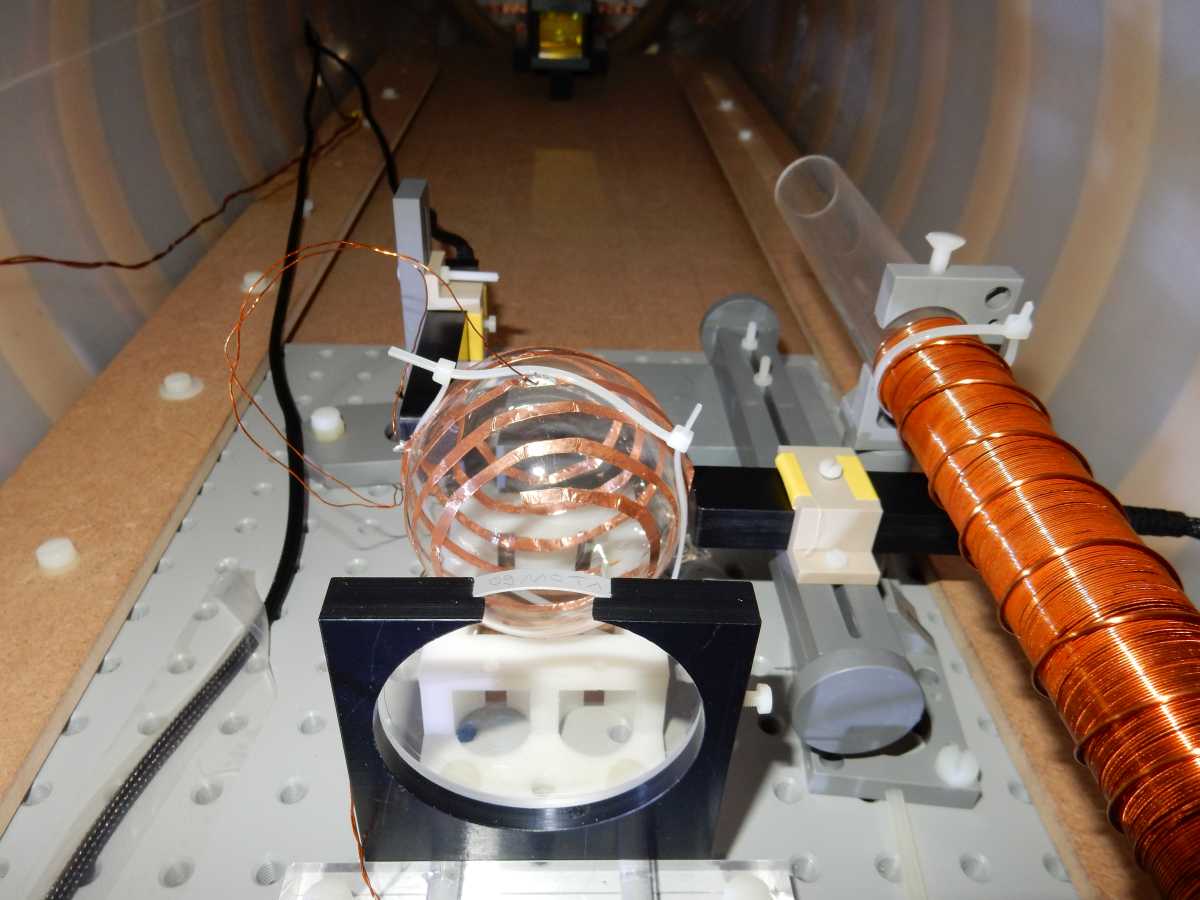}
\caption{\label{fig:appPic}Photograph of the \Rb{} and \tHe{} magnetometers being installed into the shield. The Tesla coil on the right was used to ignite a discharge in the \tHe{} cell to enable MEOP.}
\end{figure}

% What are the QuSpin sensors?
The hybrid magnetometer is schematically shown in Fig.~\ref{fig:apparatus}, and a photograph in Fig.~\ref{fig:appPic}.
Two commercially available \Rb{} zero-field magnetometers\footnote{QuSpin QZFM~\cite{Osborne2018-fw}} were used for the measurements to observe the spin precession of the \tHe{} inside the glass cell.
Each \Rb{} magnetometer unit contains a \unit[6]{mm} cubic rubidium cell, through which a \unit[795]{nm} laser tuned to the D1 line of \Rb{} is passed for optical pumping and detection~\cite{Osborne2018-fw}. A \unit[923]{Hz} oscillating magnetic field is applied to the atoms in the cell. Two separate orthogonal magnetic fields (with a \unit[90]{$^\circ$} phase shift) are applied across the cell, which, when demodulated, enables two sensitive axes to be measured. The magnetometer has a dynamic range of \unit[$\pm$5]{nT}.
Use of the sensor in two-axis mode reduces the sensitivity by around 30\%.
The internal correction coils of the \Rb{} magnetometers were not used, as the local DC fields affect the residual field value at the \tHe{} cell. \par
%
% 3He cell and the magnetometer arrangement
In this experiment a \unit[70]{mm} diameter \unit[1]{mbar} \tHe{} cell was used.
If the \tHe{} was 100\% polarised, then
a \Rb{} cell located \unit[39]{mm} from the centre of the %our 100\% polarised 
\tHe{} cell, orientated 45$^{\circ}$ from the initial polarisation direction, would measure a field of \unit[80]{pT} along the $Z$ axis.
% Polarisation optics, mirror and laser
The \tHe{} was polarised by optical pumping of the gas using circularly polarised \unit[1083]{nm} light from a commercially available \unit[5]{W} ytterbium doped fibre laser\footnote{Keopsys CYFL-GIGA}.
The QuSpin sensors were orientated to enable at least one reading in each of three axes.
%
% Shield and coils
The hybrid magnetometer was placed in a 5-layer $\mu$-metal shield, allowing external magnetic noise to be suppressed by a factor $\sim 10^5$. The shield housed six coils enabling independent control for all three field components, and first-order gradients.
%
% How long polarising, mention meop and discharge.
The \tHe{} is polarised using the MEOP process, during which a discharge is struck inside the cell.
After polarisation the discharge is switched off and the laser ramped down before turning off.
% No inner coils in the Rb magnetometers
% LabVIEW controlled constant voltage supply, room temperature stabilised. therefore current in resistors
After the \tHe{} has been polarised the \Rb{} magnetometers probe the precessing \tHe{} dipole. \par
The entire experiment was located in a closed area with a controlled temperature stability of \unit[$\pm$1]{$^{\circ}$C}.
A data acquisition system recorded the outputs from the QuSpin electronics module with a \unit[200]{Hz} sampling rate.  The data collection device was interfaced with a constant current supply, simultaneously enabling control of the current in three correction coils.
%
%%%%%%%%%%%%%%%%%%%%%%%%%%%%%%%%%%%%%%%%%%%%%%%%%%%%%%%%%%
\section{Field minimisation}
%%%%%%%%%%%%%%%%%%%%%%%%%%%%%%%%%%%%%%%%%%%%%%%%%%%%%%%%%%%%%%
\subsection{Method} \label{sec:minField}
The spin precession frequency $f_{\mathrm{\tHe{}}}$ can be converted to an absolute field at the \tHe{} cell position by
\begin{equation}
    |B_{He}| = f_{He} \frac{2\pi}{\gamma_{He}},
    \label{eqn:gyroRatio}
\end{equation}
where \unit[$\gamma_{He}/2\pi = 32.43410084(81)$]{Hz/$\mu$T}~\cite{Mohr2015-vh} is the gyromagnetic ratio of \tHe{}.
The field measured by the \tHe{} cell inside the $\mu$-metal shield is the absolute value of the residual field $B_R$, which can be expressed by the orthogonal axis components according to
\begin{equation}
|B_R| = \sqrt{B_{Rx}^2 + B_{Ry}^2 + B_{Rz}^2}.
\end{equation}
The shield was equipped with coils which produce additional fields along the different axes. Assuming perfectly orientated coils and neglecting inhomogeneities of the produced fields, the residual field together with the coil contribution can be expressed as
\begin{equation}
    |B_{R+C}|(I_x,I_y,I_z) = \sqrt{(\mathcal{B}_{x})^2+(\mathcal{B}_{y})^2+(\mathcal{B}_{z})^2},
    \label{eqn:scanX}
\end{equation}
where $\mathcal{B}_{i} = B_{Ri}-A_iI_i$, with $A$ and $I$ the coil coefficients and the currents applied to the coils in the orthogonal directions, respectively, and $i=x,y,z$. Each summand under the square root can be set to zero independently by adjusting the corresponding current; this can easily be done by minimising the function $|B_{R+C}|(I_x,I_y,I_z)$ sequentially.
After adjusting $I_x$, $I_y$ and $I_z$, $|B_{R+C}|$ should be zero if the coils are perfectly aligned and the \tHe{} cell is small in diameter so that gradient effects can be neglected. As long as Eq.~\ref{eqn:scanX} is valid, and if the coil coefficients $A_x$, $A_y$ and $A_z$ are measured in advance, the components of the residual field can be calculated by
\begin{equation}
    B_{Ri} = A_iI_{i,optimum}.
    \label{eqn:currentComponent}
\end{equation}
%
%where $i = x, y, z$. 
\par

For the more general case including some misalignment or inhomogeneity in the fields generated by the coils, Eq.~\ref{eqn:scanX} can be written as
\begin{equation}
   |\vec{B}_{R+C}|=\sqrt{ \sum_{i=1}^3( B_{Ri} - \sum_{j=1}^3 A_{ij}I_j )^2 },
    \label{eqn:sumIJ}
\end{equation}
where $A_{ij}$ is now a matrix with  $A_x$ now denoted as $A_{xx}$, $A_y$ as $A_{yy}$, and $A_z$ as $A_{zz}$.
As long as the field vectors generated by different coils are not linearly dependent, the matrix $A_{ij}$ should be invertible, and there will be exactly one set of currents that will compensate a particular residual field. Eq.~\ref{eqn:sumIJ} can  be minimised by sequentially adjusting each coil current to find the minimum field point, which if the vectors that make up $A_{ij}$ are not mutually orthogonal will not reach zero field after a single iteration of the three currents. Further iterations should allow further reduction in $|B_{R+C}|$, and in principle should eventually converge to zero field, although in practice this could be very time consuming. Alternatively, if the full $A_{ij}$ matrix can be determined in advance then this might allow a more clever and quicker optimisation method to be found. \par

The accuracy of the measurements strongly depends on the ability to determine the frequency of the spin precession from the QuSpin signals.
In the experiment, a mean frequency reading was taken from all the QuSpin channels with an appropriate signal-to-noise ratio.
With two QuSpin detectors, each containing two channels perpendicular to each other, one field direction is measured twice.
When precessing, the \tHe{} magnetisation precesses around $B_R$. If the initial polarisation ($\vec{k}_{He}$) happens to be in the same direction as $B_R$, no precession will occur.
The QuSpin devices are least sensitive along the direction of $B_R$, resulting in the hybrid magnetometer being arranged to measure three orthogonal axes.
%As the \tHe{} does not precess in the field $\vec{B_R} = (0,0,B_R)$, if the field components are sequentially minimised, the $Z$ component should be minimised first as the signal amplitude approaches zero when the residual field direction gets closer to the $Z$-axis.
%
If the field components are optimised sequentially the $Z$ component should be optimised first as the signal amplitude approaches zero when $\vec{B_R} \rightarrow (0,0,B_R) $.
In future measurements it would be advantageous to place the two QuSpin detectors at $+X$ and $-X$, orientated in such a way that the orthogonal channels can also measure $Y$ and $Z$. In this orientation the two opposing $X$ channels can produce a gradiometer signal, improving the signal-to-noise ratio of the \tHe{} precession by the reduction of common-mode noise~\cite{Bison2003-vz}. \par
\par
The principal limitation to the measurements is the long precession period of \tHe{} at near-zero fields.
Although the measurements are fundamentally limited by the finite $T_2$ time of the \tHe{} polarisation, if we limit our measurement time for one reading to 1 hour, the absolute value of the residual field cannot be lower than \unit[10]{pT} according to Eq.~\ref{eqn:gyroRatio}.
In the time it takes to record a single precession at these field values, other systematic effects can dominate the reading, such as the time stability of the QuSpin readings and the field inside the shield.
Also, as the Earth's frame rotates around the freely-precessing \tHe{} at a rate of an inverse sidereal day, the \Rb{} magnetometers will measure a shift in frequency up to $|\Delta f_{He}| \approx 12$~\unit{$\mu$Hz} resulting in \unit[$|\Delta B_R| \approx 357$]{fT}.

\subsection{Results and discussion}
%%%%%%%%%%%%%%%%%%%%%%%%%%%%%%%%%%%%%%%%%%%%%%%%%%%%%%%%%%%%%%
All data was taken during May 2018 at the Paul Scherrer Institute (PSI) in Villigen, Switzerland  \par
\begin{figure}
\centering
\includegraphics[width=0.9\columnwidth]{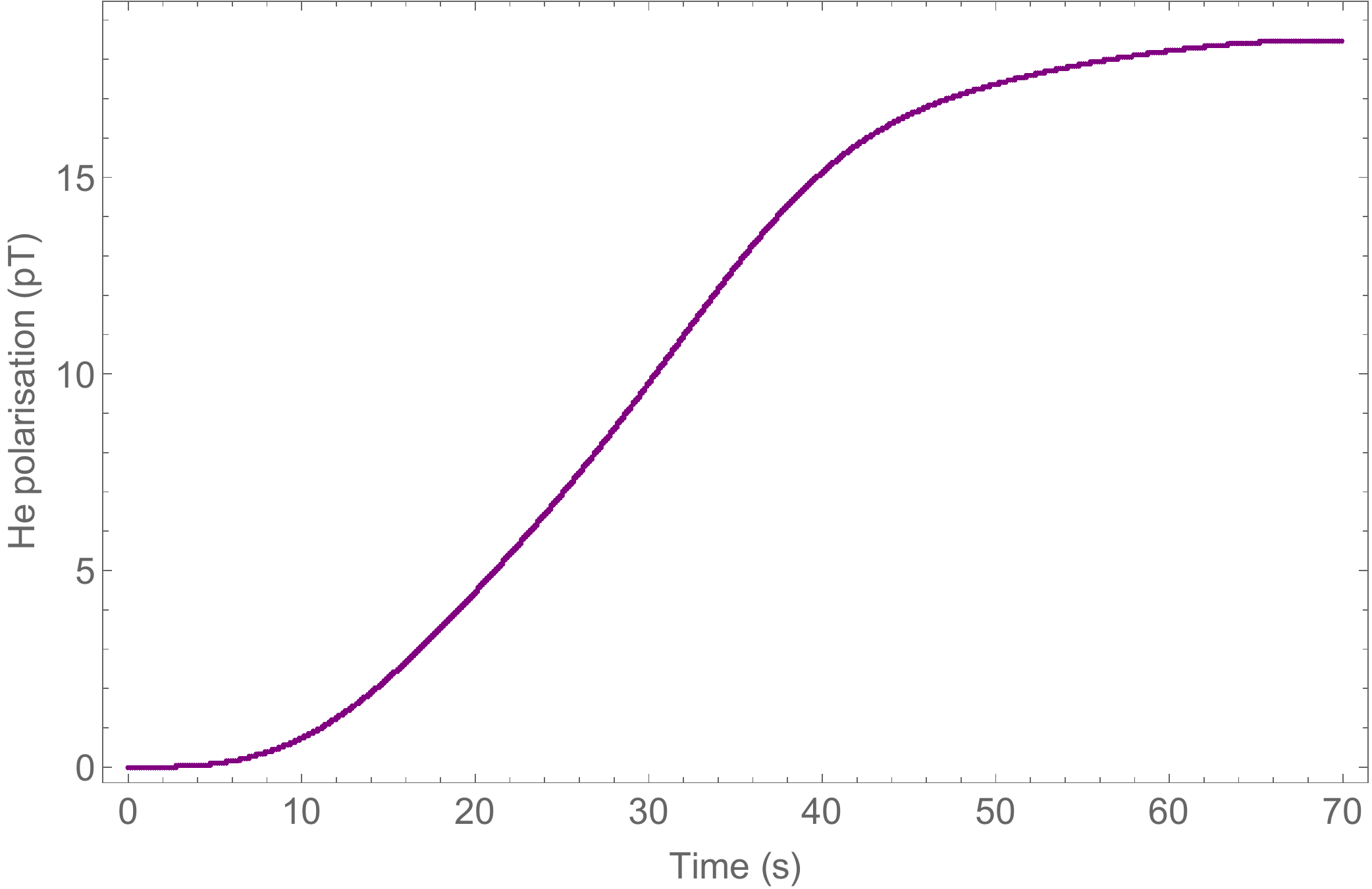}
\caption{\label{fig:hePol}
A plot displaying the field in the $Z$ direction as measured by a \Rb{} magnetometer during the \tHe{} polarisation procedure. The location of the \Rb{} magnetometer corresponds to the geometry described in Sec.~\ref{sec:expApp}, from which the \tHe{} polarisation can be approximated at 24\%.}
\end{figure}
\begin{figure}
\centering
\includegraphics[width=0.9\columnwidth]{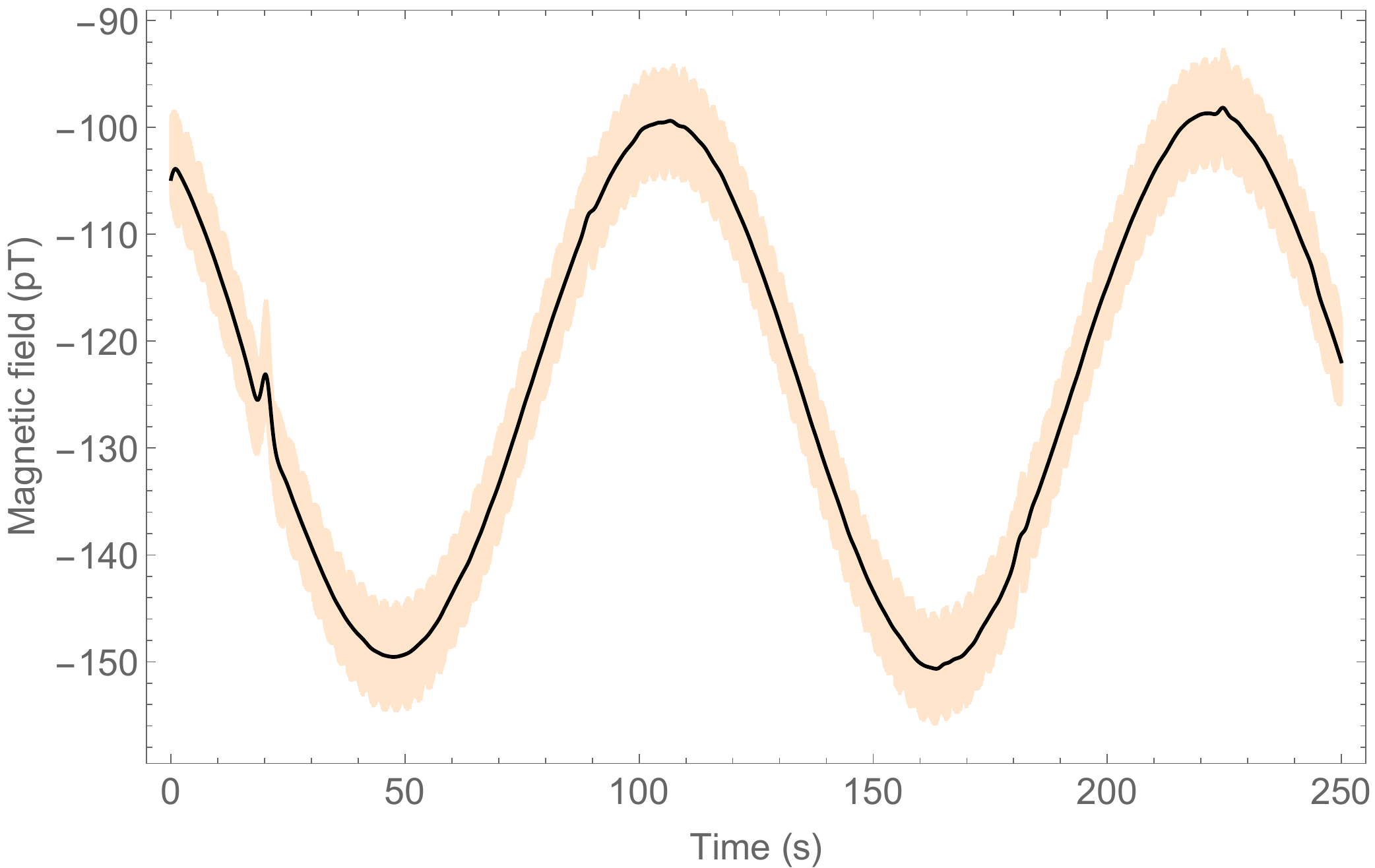}
\caption{\label{fig:rawFilt}Raw \Rb{} data displaying the oscillating \tHe{} signal and \unit[1]{Hz} low-pass filtered data (black). The field shifts at \unit[25 and 230]{s} were caused by small mechanical disturbances of the magnetic shielding. The precession frequency corresponds to a field reading of \unit[264.9]{pT}.}
\end{figure}

Figure~\ref{fig:hePol} shows the magnetisation build up in the \tHe{} cell as a function of time in the magnetic shielding as measured by the \Rb{} magnetometers.
We first measured the magnitude of the post-degauss residual field inside the shield, a plot of which can be seen in Fig.~\ref{fig:rawFilt}. By using the gyromagnetic ratio of \tHe{} the fitted frequency corresponds to a residual magnetic field of \unit[264.9]{pT}. The raw data was filtered via a \unit[1]{Hz} low-pass filter, then a sine fit was applied to the filtered data. \par
Data was taken to determine the effectiveness of the minimisation technique in reducing residual fields inside the shield, after the field had been minimised along all three field components. After the three corresponding field currents were applied to the correction coils, the true field magnitude at the predicted minima was recorded, thereby offering a comparison to the predicted minimum field from the fitting routine, and the field magnitude proper. \par
\begin{figure}
\centering
\includegraphics[width=0.9\columnwidth]{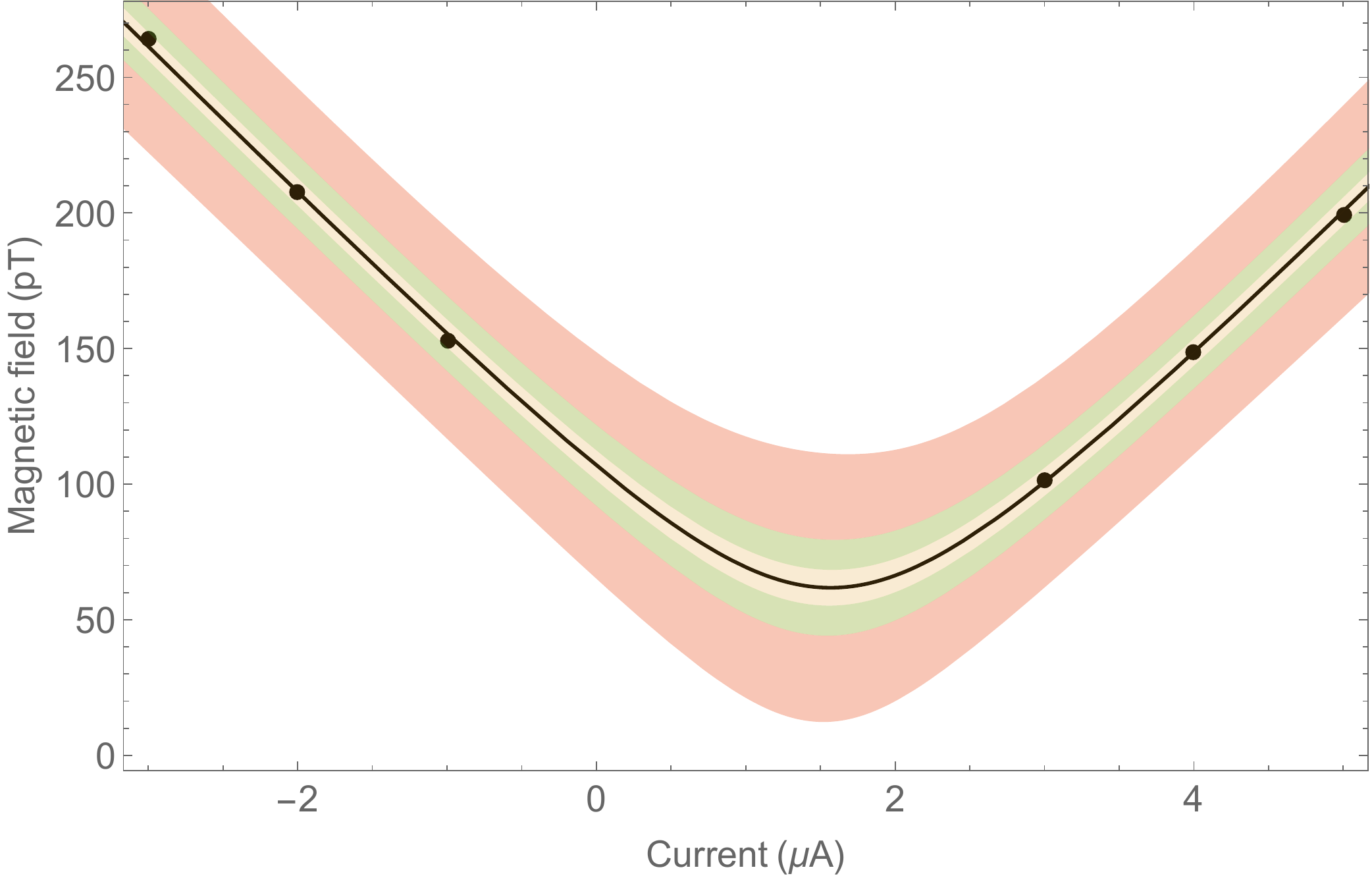}
\caption{\label{fig:xMinPlot}Scan along $X$ fitted with Eq.~\ref{eqn:scanX}. The bands correspond to the 1, 2 and 3 $\sigma$ levels. The scan was performed after the $Y$ component of the field had already been minimised.}
\end{figure}
\begin{figure}
\centering
\includegraphics[width=0.9\columnwidth]{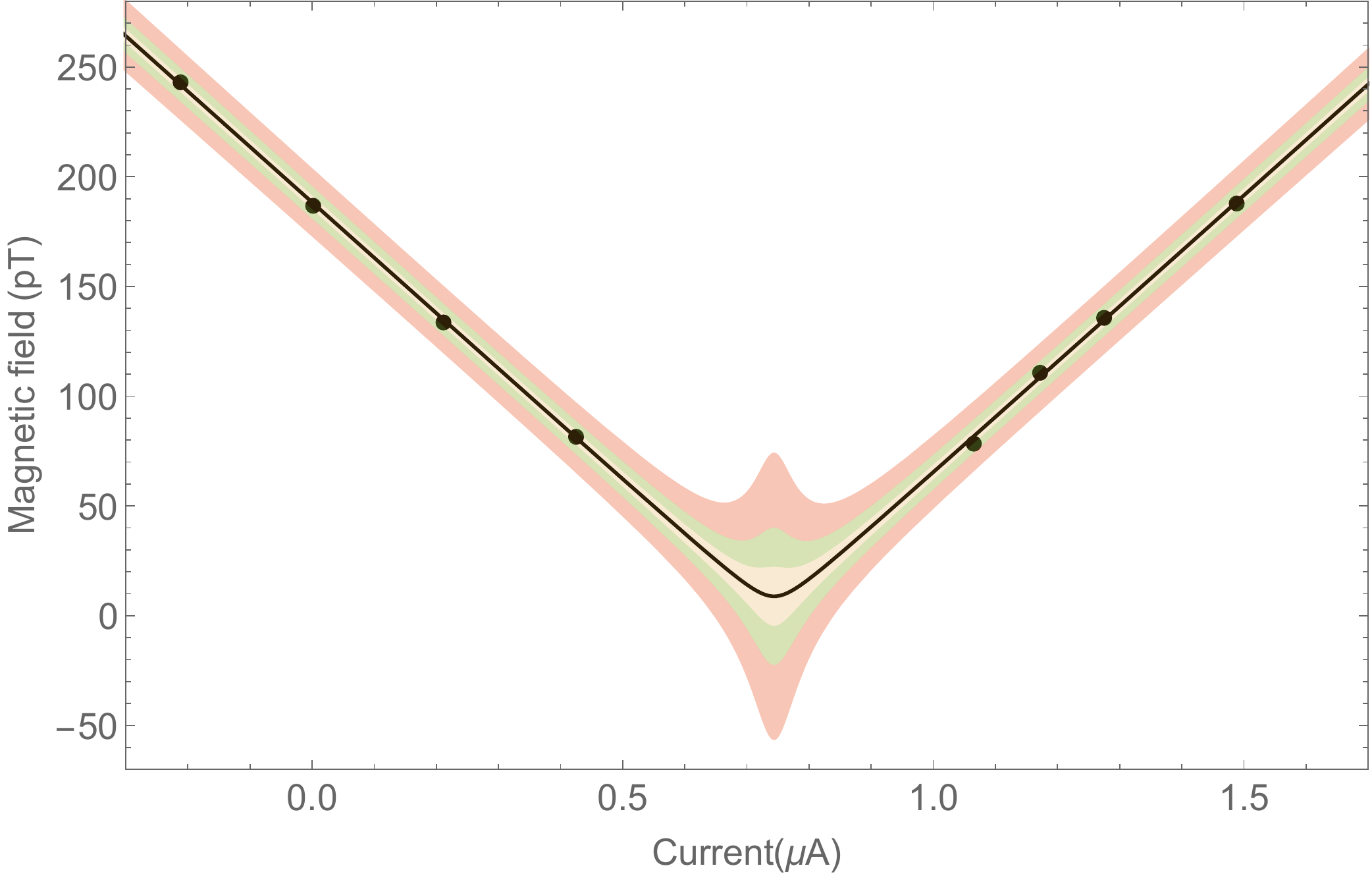}
\caption{\label{fig:fitPlot}A scan along $Z$ fitted with Eq.~\ref{eqn:scanX} immediately after the minimum current from Figure~\ref{fig:xMinPlot} was applied.}
\end{figure}

Figures~\ref{fig:xMinPlot} and~\ref{fig:fitPlot} show the results of a field minimisation procedure along $X$, followed by $Z$, after a correction current had already been applied along $Y$. Field measurements during the scan of each coil are purposefully taken displaced from the expected minimum in order to avoid excessively long precession periods. Following the scan along $X$, the field minima was estimated to be \unit[62(4)]{pT} at a current of \unit[1.57(2)]{$\mu$A}.
By scanning the current along $Z$ immediately after applying the correction current in $X$, the magnitude of the 3-axis minimised field was predicted to be \unit[9(12)]{pT} with the final correction current of \unit[0.74(2)]{$\mu$A}.
Both of the predicted field minima were calculated using Eq.~\ref{eqn:scanX} for the associated field component.
Data taken after the current in the $Z$ correction coil had been set to \unit[0.74]{$\mu$A} determined the field to be \unit[14(1)]{pT}. \par
\begin{figure}
\centering
\includegraphics[width=0.9\columnwidth]{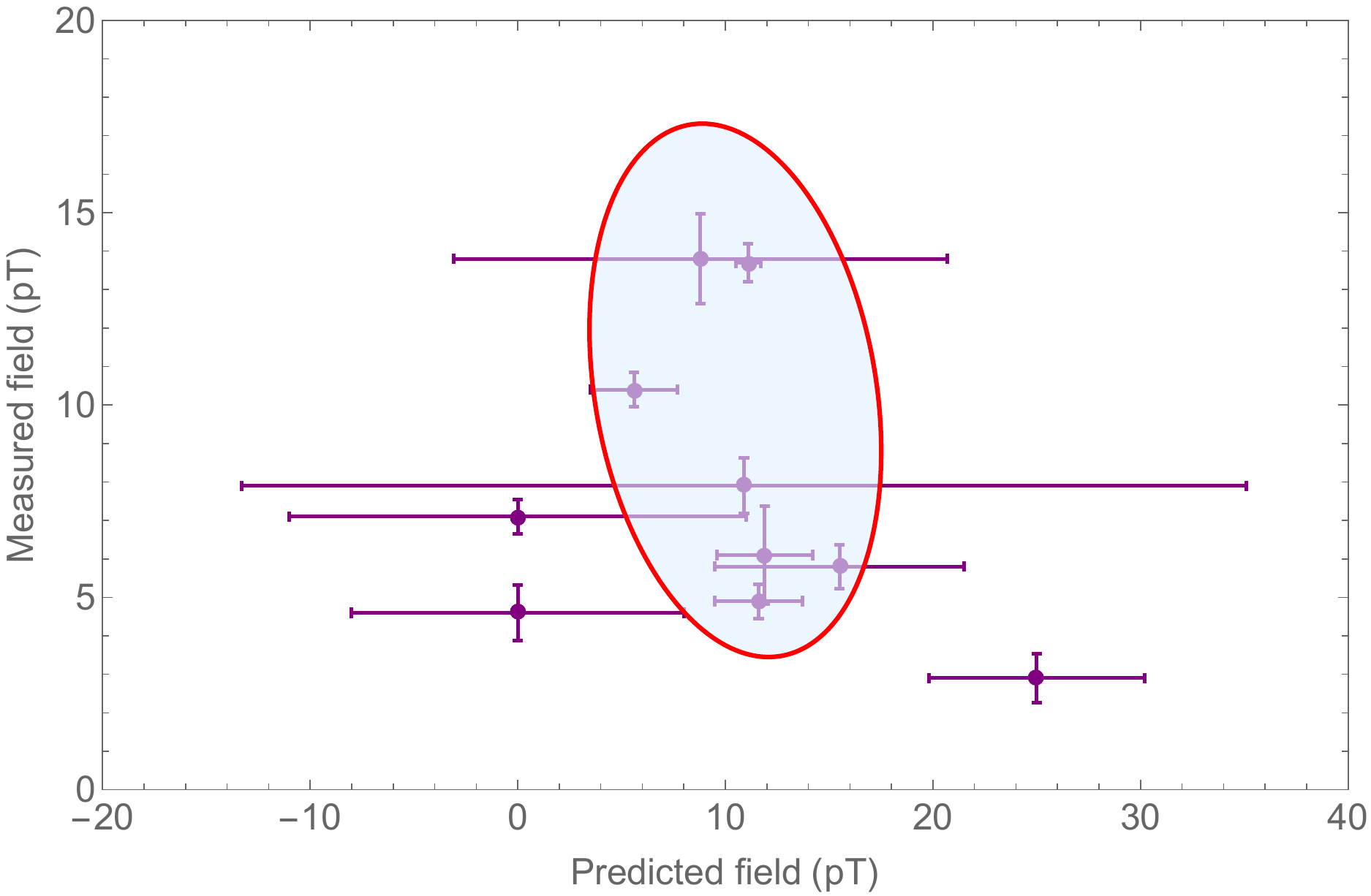}
\caption{\label{fig:minResults}Results of the minimisation measurements using the curve fit (Eq.~\ref{eqn:scanX}), along with the $1\sigma$ region.}
\end{figure}

Ten such measurements were taken during the campaign and the results can be seen in Fig.~\ref{fig:minResults}. From the plot, it can be seen that the absolute field attainable from carrying out the minimisation procedure using the hybrid magnetometer is of the order \unit[10]{pT}, with a predicted field magnitude also around \unit[10]{pT}. \par
After minimising the residual field through iteratively determining the three correction currents, it is possible to repeat the process to reach a lower residual field, reducing the coupling terms in Eq.~\ref{eqn:sumIJ}. This process strongly depends on the magnetic field stability.
%
%%%%%%%%%%%%%%%%%%%%%%%%%%%%%%%%%%%
\section{Gradient minimisation}
\begin{figure}
\centering
\includegraphics[width=0.9\columnwidth]{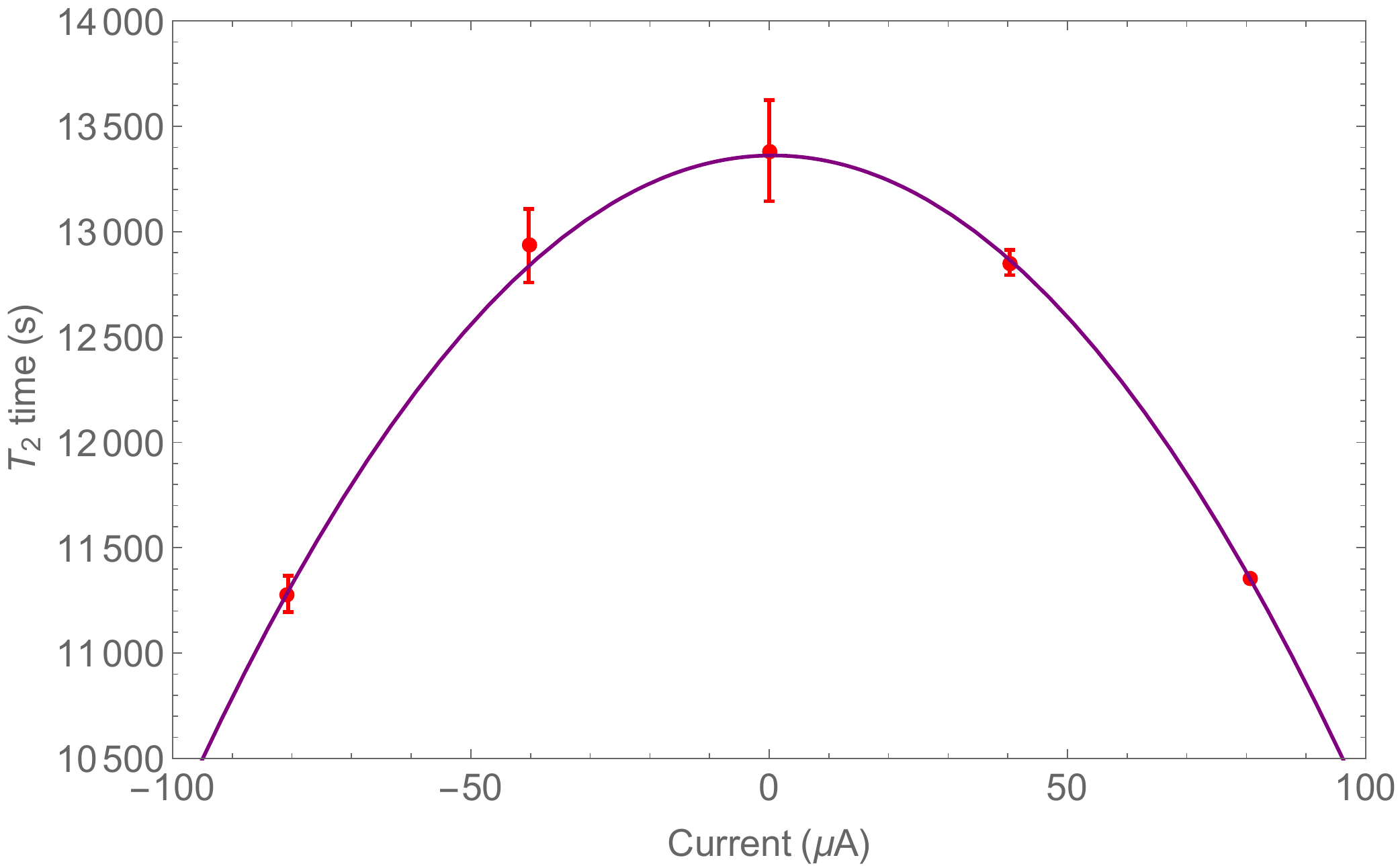}
\caption{\label{fig:T2Results}Results of the $T_2$ measurement with peak $T_2^*$ time of \unit[13\,360(30)]{s}.}
\end{figure}

%%%%%%%%%%%%%%%%%%%%%%%%%%%%%%%%%%%
The effective $T_2$ decay time ($T_2^*$) of the \tHe{} precession signal is dependent on linear field gradients through the sample volume~\cite{Cates1988-fj,Allmendinger2017-tw}.
As the field gradient is changed through the measurement volume, by use of a set of gradient coils, the fit function 
\begin{equation}
    T_2^* = A(I-I_0)^2+T_{2,peak},
\end{equation}
can be applied to determine the peak $T_2^*$ time, $T_{2,peak}$, along with the correction current $I_0$.
To reduce the initial $B_0$ amplitude, the field was first minimised along two axes following the procedure set out in Section~\ref{sec:minField}. The DAQ was instructed to apply five different currents to the $X$ gradient coil, with two hours of data being collected at each value.
The data from each current was analysed to extract the $T_2^*$ time. The results of the measurement are shown in Fig.~\ref{fig:T2Results}. The results show that when the field gradient along $X$ is minimised with an applied current of \unit[0.6(4)]{$\mu$A}, the peak $T_2^*$ time has a value of \unit[13\,360(30)]{s}.
If the optimum $T_2$ time is known, this value can be used to set limits on the remaining field gradients.
%
%%%%%%%%%%%%%%%%%%%%%%%%%%%%%%%%%%%%%%%%%%%%%%%%%%
\section{Conclusion}
%%%%%%%%%%%%%%%%%%%%%%%%%%%%%%%%%%%%%%%%%%%%%%
We have demonstrated the process of using \Rb{} zero-field magnetometers to probe the nearby precession of polarised \tHe{}, determining absolute low-field readings below 5 pT.
It was shown that the minimisation procedure utilised in the 5-layer test magnetic shield at PSI results in a consistently achievable internal magnetic field at the \unit[10]{pT} level.
The results, a factor of 30 above the sensitivity limit from the Earth's rotation, were generally limited by the field stability inside the magnetic shielding. The results could further be improved by the use of an external field compensation system \cite{Afach2014-ei} around the shielding.
Further measurements demonstrated the ability to also minimise magnetic field gradients through the volume of the \tHe{} cell.
By providing absolute field readings in the pT range, 
the results presented here offer a new method for %absolute field readings in the pT range, 
optimising the magnetic field environment in a degaussed shield,
by measuring residual magnetic fields and determining the relevant correction coil currents to cancel them.
This then provides a near zero-field base onto which any desired magnetic field configuration can be applied using an appropriately designed field coil. We note that the ultimate stability and precision of the resulting field will also depend strongly on the current supplies used, and influence of the surrounding magnetic shielding.
The \tHe{}/\Rb{} hybrid magnetometer can then be used to also provide an absolute calibration for the applied field configuration if fields below 5 nT are used, or the system described in \cite{Koch2015-vq} could be used for larger fields. In the present work we are minimising the volume averaged residual field in a 70~mm diameter \tHe{} cell. Larger zero-field regions could in principle be accomplished if the position of the \tHe{}/\Rb{} magnetometer was scannable within the shield, or by using an array of such sensors. This method can allow greatly improved understanding of the field configuration in shielded environments, providing a very useful tool for precision measurements. 

%has the potential to be used in upcoming nEDM experiments, by measuring post-degaussed fields to identify anomalies, if the apparatus were able to be moved around the shielded volume.
%The magnetic field measurement device can be used for the cancellation of both zeroth and first-order magnetic field gradients.
%
\section*{Author contribution statement}
C. Abel and G. Bison conceived of the presented idea.
C. Abel, G. Bison and W. C. Griffith. planned, collected and analysed the data.
K. Kirch, B. Lauss and A. Mtchedlishvili conceived and constructed the magnetic shield.
H. C. Koch and M. Pototschnig constructed the coil system.
K. Kirch, B. Lauss and P. Schmidt-Wellenburg conceived and constructed the temperature stabilised environment.
All authors discussed the results and contributed to the final manuscript with large inputs from C. Abel, G. Bison, W. C. Griffith, W. Heil, A. Schnabel, and J. Voigt.
\par
This work was funded in part by the U.K. Science and Technology Facilities Council (STFC) through Grants No. ST/N000307/1 and No. ST/M503836/1, as well as by the School of Mathematical and Physical Sciences at the University of Sussex.
We acknowledge the support of the Core Facility 'Metrology of Ultra-Low Magnetic Fields' at Physikalisch-Technische Bundesanstalt which receives funding from the Deutsche Forschungsgemeinschaft – (KO 5321/3-1 and TR 408/11-1), and support for the apparatus at PSI aided through grants from the Deutsche Forschungsgemeinschaft (BI 1424/2-1, BI 1424/3-1) and the Swiss National Science Foundation (163988, 172626).
% BibTeX users please use
\bibliographystyle{epj}

\end{document}